\documentclass[10 pt]{article}
\usepackage{graphicx}
\usepackage{epsfig}
\usepackage{color}

\usepackage{amssymb,amsmath}
\usepackage{graphicx}  
\newcommand{\beq}{\begin{equation}}
\newcommand{\eeq}{\end{equation}}
\newcommand{\be}{\begin{equation}}
\newcommand{\ee}{\end{equation}}
\newcommand{\bea}{\begin{eqnarray}}
\newcommand{\eea}{\end{eqnarray}}

\fontencoding{T1}
\fontfamily{garamond}
\fontseries{m}
\fontshape{it}
\fontsize{13}{15}
\selectfont

\begin{document}

\title{Indications of a Four-Quark Structure for the $X(3872)$ and
    $X(3876)$ Particles from Recent Belle and BABAR Data}
\author{L.~Maiani$^a$, A.D. Polosa$^b$, V.~Riquer$^b$, \\
$^a$Dip. Fisica, Universit\`a  di Roma ``La Sapienza'' and INFN, Roma, Italy\\
$^b$INFN, Sezione di Roma, Roma, Italy}
\date{\today}
\maketitle
\begin{abstract}
Recent results by BELLE and BaBar point to the existence of a second $X$ particle decaying in $D^0{\bar D}^0\pi^0$, a few MeV above the $X(3872)$.
We identify the two X states with the neutral particles predicted by the 4-quark model and show that production and decays are consistent with this assignment. We consider the  yet-to-be-observed charged partners and give new hints on how to look for them.
\end{abstract}


The firmest prediction of the four-quark model of X(3872)~~\cite{MPPR} is the existence of a complex of four related states, two neutrals and two charged:
\bea
&& X_u=[cu][{\bar c}{\bar u}];~~~X_d=[cd][{\bar c}{\bar d}];\notag \\
&& X^+=[cu][{\bar c}{\bar d}];~~~X^-=[cd][{\bar c}{\bar u}]
\label{Xstates}
\eea
with masses differing by few-to several MeV, i.e. of the order of the electromagnetic and of the $u-d$ quark mass differences.

Indications of a mass difference between the X state decaying in $ J/\psi \pi^+ \pi^-$~\cite{X1} and the one decaying in $D^0{\bar D}^0\pi^0$ have been reported by Belle and BaBar:
\bea
&&M( J/\psi \pi^+ \pi^-)= 3871.2 \pm 0.5~{\rm MeV}~~~~{\rm (World~Average~\cite{PDG})}
 \label{wordave}
\\
&& M( D^0{\bar D}^0\pi^0)=\left\{ \begin{array}{c} 3875.4 \pm 0.7^{+1.2}_{-2.0}~{\rm MeV}~{\rm (Belle\cite{BelleDm})}\\
3875.6  \pm 0.7^{+1.4}_{-1.5}~{\rm MeV}~{\rm (BaBar\cite{BBDm})}       \end{array}\right .
\label{newmass}
\eea

If confirmed, these results would indicate the existence of two different neutral states, X(3872) and X(3876), with a mass difference of the order expected for $X_u$ and $X_d$.
The available information on the production rates of these two states in $B^+$ and $B^0$ decays is summarized in Table~\ref{tab1}.

In this note, we identify the states (\ref{wordave}) and (\ref{newmass}) with the neutral states in (\ref{Xstates}), by considering a scheme whereby $X_u$ and $X_d$ are both generically produced in $B^+$ and $B^0$ decays, but then decay predominantly in disjoint channels.
A closer analysis of the decay rates, reported below, leads us to the identification:
\bea
&&X_u={\rm X~state~decaying~into}~D^0{\bar D}^0\pi^0~=~X(3876) \notag \\
&&X_d={\rm X~state~decaying~into}~J/\psi \pi^+\pi^-=X(3872)
\label{assign}
\eea
Data are still scarce so, not to complicate the analysis, we neglect possible $X_{u,d}$ mixing~\cite{MPPR}.

\begin{table}[htb]
\begin{center}
\caption{{\footnotesize Summary of the available information on production and decay rates of X states in $B^+$ and $B^0$ decays. For each entry, we give BELLE data and BaBar data in the first and second line, respectively. The Belle $B^0/B^+$ ratio is computed from individual rates with errors added in quadrature.}}
\label{tab1}
\vskip0.5cm
\begin{tabular}{@{}l|c|c}
~~~~  & $f= J/\psi\pi^+\pi^-$ &\footnotesize $f= D^0{\bar D}^0\pi^0$ \\ 
\hline
${\cal B}(B^\pm \to K^\pm X){\cal B}(X \to f)\cdot$10$^5$&
$
\begin{array}{c} 1.05 \pm 0.18\cite{X1}\\ 1.01\pm 0.25 \pm 0.10\cite{BBDm0}\end{array}
$ 
& $
\begin{array}{c} 10.7 \pm 3.1^{1.9}_{3.3}~\cite{BelleDm}\\ ----\end{array}
$ \\
\hline
${\cal B}(B^0 \to K^0 X){\cal B}(X \to f)\cdot$10$^5$&
$
\begin{array}{c} ----\\ 0.51 \pm 0.28 \pm 0.07~\cite{BBDm0}\end{array}
$ &
$
\begin{array}{c} 17.3 \pm 7.0 ^{3.1}_{5.3}~\cite{BelleDm}\\----
\end{array}$\\ 
\hline
($B^0/B^+$)$_{f}$ & 
$
\begin{array}{c} ----\\0.50 \pm 0.30 \pm 0.05~\cite{BBDm0} \end{array}
$ &
$
\begin{array}{c} 1.62 \pm 0.80 \\ 2.23 \pm 0.93 \pm 0.55~\cite{BBDm}\end{array}
$ \\
\hline
\end{tabular}\\[2pt]
\end{center}
\end{table}

We show that the identification (\ref{assign}) leads to a simple relation for the ratios of $B^0$ to $B^+$ rates in the two final states:
\be
\left(\frac{B^0}{B^+}\right)_{J/\psi}=\left[\left(\frac{B^0}{B^+}\right)_{D\bar D\pi}\right]^{-1}
\label{relation1}
\ee
The data suggest that Eq.~(\ref{relation1}) is satisfied, albeit with still rather large errors. Moreover, we argue that the decay rates expected for tetraquark states satisfy naturally the relations:
\bea
&&\Gamma(X_u \to D^0 \bar D^0 \pi^0)>> \Gamma(X_u \to J/\psi\pi^+ \pi^-)\simeq \notag \\
&&\simeq \Gamma(X_d \to J/\psi\pi^+ \pi^-)>>\Gamma(X_d \to D^0 \bar D^0 \pi^0)
\label{ordering}
\eea
which explain why X states are seen in different channels and also account for the relative size of the branching ratios implied by Table~\ref{tab1}. Finally, we comment on the non-observation of $X^\pm$ in B decays. An improvement   by a factor of five on present  bounds would test conclusively our scheme and a search for  $X^+\to D^+ \bar D^0 \pi^0$ would  provide very useful information.

The only seemingly counter intuitive feature of the assignment (\ref{assign}) is the ordering of $X_u$ and $X_d$ masses. Being the down quark heavier than the up quark, we could expect $X_d$ to be the heaviest particle. However, this argument neglects the electrostatic interactions inside the diquarks, which would be repulsive for $X_u$ and attractive for $X_d$ and could  change the ordering of the states for configurations with closely packed diquarks. At the moment we are unable to go beyond this qualitative statement and have to defer the problem to further study.

The absence of a visible structure in the mass distribution of $J/\psi \pi^+ \pi^-$~\cite{X1} led us previously to suggest~\cite{MPPR} an interference of the weak amplitudes in B decays such that one of the neutral mass states (\ref{Xstates})  would appear  in $B^+$ and the other in $B^0$ decays. This prediction has not been supported by subsequent studies that showed that the states decaying in $D^0{\bar D}^0\pi^0$ in both $B^0$ and $B^+$ have the same mass within errors~\cite{BBDm}. 
The situation is less clear for the state decaying in $J/\psi \pi^+\pi^-$~\cite{BBDm0}, but still not incompatible with the states in $B^0$ and $B^+$ decay being the same one. However, the assumption made implicitly in~\cite{MPPR}, that  $X_u$ and $X_d$ would decay in  $J/\psi \pi^+ \pi^-$ with similar branching ratios is not justified and the earlier scheme is superseded by the one presented here.

We analyze B non leptonic decay amplitudes under the hypothesis (\ref{assign}). In quark language, the allowed weak transition for $B^+$, is:
\bea
{\bar b}+(u) \to {\bar c} + c{\bar s}  +(u)+ q{\bar q}
\label{weakT}
\eea
(spectator quark in parenthesis). We allow for the creation of an additional $q\bar q$ pair from vacuum, to obtain the $K^+ +X$ state. There are two independent amplitudes~\cite{MPPR}, which correspond to make the $K^+$ by joining 
$\bar s$ either to the spectator or to the u quark from the sea. Denoting by A and B the two amplitudes, one finds:
\bea
&&{\cal A}(B^+\to K^+ X_u)=A+B= {\cal A}(B^0\to K^0 X_d)\notag \\
&&{\cal A}(B^+\to K^+ X_d)=A={\cal A}(B^0\to K^0 X_u)
\label{weakA}\\
&&{\cal A}(B^+\to K^0 X^+)=B ={\cal A}(B^0\to K^+ X^-)
\label{weakXch}
\eea
Although derived from the quark picture~(\ref{weakT}), the previous structure  follows generally from the isospin of  the weak Hamiltonian and of the $X_{u,d}$ states. Indeed,  $B={\cal A}(B\to X_{I=1}K)$ and, as such, this parameter  controls the production of $X^\pm$.  Isospin symmetry relates $B^+$ to $B^0$ decays. One easily finds the result anticipated in Eq.~(\ref{relation1}):
\bea
&&\left(\frac{B^0}{B^+}\right)_{J/\psi}=
\frac{{\cal B}(B^0\to K^0 X_d){\cal B}(X_d\to J/\psi\pi^+\pi^-)}{{\cal B}(B^+\to K^+ X_d){\cal B}(X_d\to J/\psi\pi^+\pi^-)}
=\frac{{\cal B}(B^0\to K^0 X_d)}{{\cal B}(B^+\to K^+ X_d)}=\notag \\
&&=\frac{{\cal B}(B^+\to K^+ X_u)}{{\cal B}(B^0\to K^0 X_u)}=\frac{{\cal B}(B^+\to K^+ X_u){\cal B}(X_u\to  D\bar D\pi)}{{\cal B}(B^0\to K^0 X_u){\cal B}(X_u\to D\bar D\pi)}=\left[\left(\frac{B^0}{B^+}\right)_{D\bar D\pi}\right]^{-1}\notag
\eea

A glance at Table~\ref{tab1} shows that Eq.~(\ref{relation1}) is satisfied by the central values of the $B^0$ to $B^+$ rate ratios. Within still large errors, the two X particles (\ref{wordave}) and (\ref{newmass}) look indeed like being related to each other by the exchange $u\leftrightarrow d$. 

The $B^0$ to $B^+$ rate ratio for one decay mode fixes the amplitude ratio B/A, up to a twofold ambiguity:
\bea
&&\left(\frac{B^0}{B^+}\right)_{D\bar D\pi}= \left|\frac{A}{A+B}\right|^2\simeq 2 \to \frac{B}{A}\simeq \left\{ \begin{array}{c}-0.3\\-1.7
\end{array}\right .
\label{solutions}
\eea

Finally, from the previous relation and Table~\ref{tab1}, we can determine the ratio of branching ratios:
\bea
&&R=\frac{{\cal B}(B^+ \to K^+ X_u){\cal B}(X_u \to D^0{\bar D}^0\pi^0)}{{\cal B}(B^+ \to K^+ X_d){\cal B}(X_d \to J/\psi \pi^+\pi^-)}= \left|\frac{A+B}{A}\right|^2~\frac{{\cal B}(X_u \to D^0{\bar D}^0\pi^0)}{{\cal B}(X_d \to J/\psi \pi^+\pi^-)}\simeq 10 \notag
\eea
that is:
\be
{\cal B}(X_d \to J/\psi \pi^+\pi^-)\simeq \frac{1}{20}{\cal B}(X_u \to D^0{\bar D}^0\pi^0)
\label{Bratio}
\ee

We turn now to consider the decay rates of $X_{u,d}$. These are expected to be made by two, non interfering, components: (i) annihilation into gluons leading to multihadron, uncharmed, states and (ii) quark rearrangement, leading to charmonium containing or to charmed pair states. 

For the quantum numbers $J^{PC}$=$1^{++}$, one needs annihilation into more than two gluons. The corresponding rate should be very similar to the decay rate of the P-wave, $1^{++}$, charmonium~\cite{PDG}:
\be
\Gamma_{ann}(X)\simeq \Gamma(\chi_{c1})=0.96~{\rm MeV}
\label{gammann}
\ee

One could also consider the annihilation of the $c{\bar c}$ pair in lowest order, leading to X$\to g g +q{\bar q}$. However, it is easily seen that the $c{\bar c}$ pair is in J=1~\cite{MPPR,Voloshin} and it cannot annihilate into two gluons. More gluons lead again to the previous estimate: 1~MeV sets the scale of the background, multihadronic, decays of $X_{u,d}$.

Quark rearrangement is the transition of a diquark-antidiquark, $[q_1 q_2]_{\bf \bar 3}[\bar q_3 \bar q_4]_{\bf 3}$, into two colorless $q\bar q$ pairs, $(\bar q_3 q_2)_{\bf 0}(q_1\bar q_4)_{\bf 0}$. 
Within the constituent quark model, we proposed~\cite{MPPRscal, MPPR}  the mechanism for such a transition to be the simultaneous tunneling of $q_1$ and $\bar q_3$ through the potential barrier that keeps the diquark and the anti-diquark separately bound. Transitions to $\psi V$, V=$\rho, \omega$, give rise to $\psi 2\pi$ and $\psi 3 \pi$ decays, transitions to ${\bar D}^*D$ and ${\bar D}D^*$ give rise to $D{\bar D}\pi$ decays. 

If annihilation is small, as indicated by the small width (\ref{gammann}), quark flavors cannot be changed, i.e. $u{\bar u}\leftrightarrow d{\bar d}$ is suppressed: {\it the prominent decay of X(3876) in $D^0 {\bar D}^{*0}$ identifies this particle as $X_u$}. Note that the successful relation (\ref{relation1}) would be consistent also with $X_u$ being the lightest state.

We describe these transitions with the effective couplings: 
\bea
&&{\cal L}_{eff}=\lambda^u_{\psi V}~\frac{1}{M_\rho}\epsilon^{\mu\nu\rho\sigma}(p_V)_\mu V_{\nu}\psi_\rho X^{(u)}_\sigma + \lambda^u_{D^* D}~X^{(u)\mu} ({\bar D}^{*0}_\mu D^0-{\bar D}^0 D^{*0}_\mu)\simeq \notag \\
&&\simeq \lambda^u_{\psi V}~{\bf X}^{(u)}\cdot ({\bf V}\times {\bf \psi} )+\lambda^u_{D^* D}~{\bf X}^{(u)}\cdot ({\bf D}^{*0}D^0-{\bf D}^{*0}{\bar D}^0)
\eea
(and similar formulae for $X_d$). In the r.h.s. we give explicitly the non-relativistic limit, for comparison with~\cite{MPPR}. Each decay is determined by  one phenomenological coupling, $\lambda$, with dimension of mass. Individual rates to the final state $f$ are written as:
\be
\Gamma(X_{u,d}\to f)=|\lambda|^2 \gamma(f)_{u,d}
\ee
The reduced rates, $\gamma(f)$, are computed from the appropriate Feynman diagrams with V or $D^*$ exchange, e.g. X$\to \psi \rho^0 \to \psi \pi^+\pi^-$. 

We report in Table~\ref{tab2} the reduced rates of representative channels. The drop of the X(3872) rate in $D^0{\bar D}^0\pi^0$ with respect to X(3876) is due to the drastic reduction of phase space. The channels $D^+ D^- \pi^0$, $D^+ {\bar D}^0\pi^-$ and $D^- D^0\pi^+$ are forbidden for X(3872) and heavily suppressed for X(3876), we report only the first one which has the largest phase space.

\begin{table}[htb]
\begin{center}
\caption{{\footnotesize Reduced rates of exclusive decays of the X states, $\gamma(f)$, in units MeV/GeV$^2$. Vector mesons are described by a relativistic Breit-Wigner, with parameters from Ref.~\cite{PDG}(masses and widths in MeV): $M_\rho$=776, $\Gamma_\rho$= 150, $M_{D^{*0}}$= 2006.7, $ \Gamma_{D^{*0}}$=0.070$, {\cal B}(D^{*0}\to D^0\pi^0)$=0.619; $M_{D^{*+}}$= 2010.0, $ \Gamma_{D^{*+}}$=0.096$, {\cal B}(D^{*+}\to D^+\pi^0)$=0.307.}}
\label{tab2}
\vskip0.5cm
\begin{tabular}{@{}l|c|c|c|c|c}
~~~~~~ & {\footnotesize $f=J/\psi\pi^+\pi^-$} &
{\footnotesize  $f=D^0{\bar D}^0\pi^0$} &
{\footnotesize $f=D^+ D^- \pi^0$}& {\footnotesize $f=J/\psi\pi^+\pi^0$} &{\footnotesize $f=D^+ {\bar D}^0 \pi^0$} \\
\hline
{\footnotesize X(3876)=$X_u \to f$} & 0.59 & 0.26&4.5$\cdot$10$^{-7}$ &---&--- \\
{\footnotesize X(3872)=$X_d \to f$} & 0.56 & 0.0102& 0 & --- &--- \\
\hline
{\footnotesize $X^+$(3877)$\to f$} & --- & ---& ---& 1.2 & 0.129  \\
{\footnotesize $X^+$(3876)$\to f$} & --- & ---& ---& 1.2 & 0.059  \\

\hline
\end{tabular}\\[2pt]
\end{center}
\end{table}

To get the pattern of $X_u$ decays, we have to consider the relation between $\lambda_{J/\psi \rho}$ and $\lambda_{D D^*}$. We make two requirements: (i) the $X_u$ signal in the $J/\psi\pi^+\pi^-$ channel should be less than 1/3 of the $X_d$ signal, since a double structure is not observed, (ii) the rate of $X_u\to D^0\bar D^0\pi^0$ should be less than the experimental resolution of about 3 MeV. Using (\ref{weakA}), we obtain:
\bea
&&\frac{{\cal B}(B^+\to K^+ X_u){\cal B}(X_u\to J/\psi \pi^+\pi^-)}{{\cal B}(B^+\to K^+ X_d){\cal B}(X_d\to J/\psi \pi^+\pi^-)}=~ \left|\frac{A+B}{A}\right|^2~ \frac{{\cal B}(X_u\to J/\psi \pi^+\pi^-)}{{\cal B}(X_d\to J/\psi \pi^+\pi^-)}\leq \frac{1}{3}
\label{nostruct} \\
&&\Gamma(X_u\to D^0\bar D^0\pi^0)=|\lambda^u_{D^* D}|^2 \gamma(D^0\bar D^0\pi^0)_u\leq 3~{\rm MeV}
\label{exptres}
\eea
Using the results in Table~\ref{tab2} together with (\ref{Bratio}), we find:
\bea
&&\left|\frac{\lambda^u_{\psi\rho}}{\lambda^u_{D^* D}}\right|^2\leq 0.017\label{lambdaratio}\\
&& \Gamma(X_u\to J/\psi\pi^+\pi^-)\simeq \Gamma(X_d\to J/\psi\pi^+\pi^-)\leq 0.1~{\rm MeV}
\label{smallwdth}
\eea

Though very speculative, the mechanism of~\cite{MPPRscal,MPPR} suggests that the transition may strongly depend upon the masses of the quarks that have to tunnel through the barrier, i.e. the c-quark in $X_{u,d} \to J/\psi\rho$, the light quark in $X_{u,d} \to D D^*$. Thus it may not be surprising that the former transition is suppressed with respect to the latter due to the heavier mass of the c quark, counterbalancing the larger reduced rate for the $J/\psi 2\pi$ mode in Table~\ref{tab2} and justifying the first line of (\ref{ordering}). 
The suppression of  $\lambda_{J/\psi \rho}$, Eq.~(\ref{lambdaratio}), brings the corresponding rate for $X_{u,d}$, Eq.~(\ref{smallwdth}), well below the pure annihilation processes, Eq.~(\ref{gammann}).

By quark flavor conservation, $X_d$ should decay in $D^+ D^{*-}+D^- D^{*+}$, which is forbidden by phase space. The transition, $X_d \to D^0 D^{*0}$, is suppressed twice because it involves $u{\bar u}\leftrightarrow d{\bar d}$ transition and because of the small reduced rate, Table~\ref{tab2}. We conclude that $J/\psi2\pi$ is the prominent quark rearrangement decay channel for $X_d$, second line of (\ref{ordering}).

Finally, we consider the yet unobserved $X^\pm$. Experimental bounds have been reported in~\cite{BB06} (90\% confidence limits):
\bea
&&{\cal B}(B^+ \to K^0 X^+){\cal B}(X^+ \to J/\psi \pi^+\pi^0)\leq 2.2 \cdot10^{-5}\notag \\
&&{\cal B}(B^0 \to K^+ X^-){\cal B}(X^- \to J/\psi \pi^-\pi^0)\leq 0.54 \cdot10^{-5}
\label{boundXch}
\eea
Using the most unfavorable solution in (\ref{solutions}), the most stringent bound is obtained from the second line of (\ref{boundXch}), to wit:
\bea
&&{\cal B}(X^+ \to J/\psi \pi^+\pi^0)\leq\left|\frac{A+B}{B}\right|^2\times \frac{0.54}{0.51}\times {\cal B}(X(3872) \to J/\psi \pi^+\pi^-)\simeq 0.25
\label{newbnds} 
\eea
where we have also used Eq.~(\ref{Bratio}).

Multihadron decays of X$^\pm$ only arise from the multigluon+$q{\bar q}$ partonic state already considered. The corresponding width may go from the J/$\psi$ width, 93 keV, up to order of 1 MeV, Eq.~(\ref{gammann}). Therefore, with $\Gamma(X^+\to J/\psi\pi^+\pi^0)=2\Gamma(X_d\to J/\psi\pi^+\pi^-)$ and the estimate (\ref{smallwdth}) the bound (\ref{newbnds}) is still not very compelling. In addition, if the mass of $X^+$ is greater than about 3876 MeV, the mode $X^+\to D^+ {\bar D}^0 \pi^0$ becomes important and the bound (\ref{newbnds}) can be satisfied even for small multihadronic rate. 

An improvement by a factor of five on the bounds (\ref{boundXch}) could test conclusively our scheme and a search for  $X^+\to D^+ {\bar D}^0 \pi^0$ is in order.

We thank Dr. Kai Yi  for a very interesting discussion. Partial support by the HELEN-ALFA project, by SLAC and PH-TH Department at CERN are gratefully acknowledged.

\end{document}